\documentclass[useAMS,usenatbib]{mnras}
\usepackage{amsmath}
\usepackage{graphicx}
\usepackage{txfonts}
\usepackage{natbib}
\usepackage{bm}

\def\apj{ApJ}
\def\apjl{ApJL}
\def\apjs{ApJ Suppl. Ser.}
\def\apss{Astroph.Sp.Sci.}
\def\aap{A\&A}

\def\araa{ARAA}
\def\mnras{MNRAS}
\def\nat{Nature}

\def\physrep{Phys. Rep.}
\def\sovast{Sov. Astron.}

\def\beq#1{\begin{equation}\label{#1}}
\def\eeq{\end{equation}}
\def\beqa#1{\begin{eqnarray}\label{#1}}
\def\eeqa{\end{eqnarray}}

\def\Eq#1{Eq.~(\ref{#1})} 

\def\comment#1{\relax}

\title[Rapidly rotating neutron star progenitors]{Rapidly rotating neutron star progenitors}
\author[Postnov et al.] {K.A. Postnov$^{1,2,3}$\thanks{E-mail: pk@sai.msu.ru},
A.G. Kuranov$^1$, D.A. Kolesnikov$^{1, 2}$, S.B. Popov$^{1}$,  N.K. Porayko$^{1, 2, 3, 4}$\\
$^1$ Sternberg Astronomical Institute, Moscow M.V. Lomonosov State University, Universitetskij pr., 13,  Moscow 119992, Russia\\
$^{2}$ Faculty of Physics,
M.V. Lomonosov Moscow State University,
Leninskie Gory, Moscow 119991, Russia\\
$^{3}$ Institute of Theoretical and Experimental Physics, Moscow, Russia\\
$^{4}$ Max-Planck-Institut f\"ur Radioastronomie, Bonn, Germany}	

\begin{document}

\date{Received ... Accepted ...}
\pagerange{\pageref{firstpage}--\pageref{lastpage}} \pubyear{2015}

\maketitle

\label{firstpage}
\begin{abstract}

Rotating proto-neutron stars can be important sources of gravitational waves to be
searched for by present-day and future interferometric detectors. It was demonstrated by
Imshennik that in extreme cases the rapid rotation of a collapsing stellar core may
lead to fission and formation of a binary proto-neutron star which subsequently merges
due to gravitational wave emission. In the present paper, we show that such dynamically
unstable collapsing stellar cores may be the product of a former merger process of two
stellar cores in a common envelope. We applied population synthesis calculations to assess the
expected fraction of such rapidly rotating stellar cores which may lead to fission and
formation of a pair of proto-neutron stars. We have used the BSE population synthesis code
supplemented with a new treatment of stellar core rotation during the evolution via
effective core-envelope coupling, characterized by the coupling time, $\tau_c$.
The validity of this approach is checked by direct MESA calculations of the
evolution of a rotating 15~$M_\odot$ star. From comparison of the calculated spin distribution of young neutron stars with
the observed one, reported by Popov and Turolla, we infer the value $\tau_c\simeq 5\times 10^5$~years.
We show that merging of stellar cores in common envelopes 
can lead to collapses with dynamically unstable proto-neutron stars, with their
formation rate being $\sim 0.1-1\%$ of the total core collapses, depending on
the common envelope efficiency.

\end{abstract}


\section{Introduction}

\label{intro}

The present-day attempts to model the explosion of core collapse supernova  (SN) in the framework of neutrino mechanism proposed 
half a century ago \citep{1966ApJ...143..626C} are almost successful in 3D (see, e.g., recent calculations \citealt{2015ApJ...808L..42M} and references therein).
Clearly, rotating neutron stars (NS) that result from core-collapse SNe and are observed as radio pulsars shows that the rotation of the collapsing core of a massive star should be taken into account in the SN explosion calculations 
(see \citealt{2012ARA&A..50..107L} for a review of 
rotational effects on the evolution of massive stars in single and binary systems).  
The rotation and magnetic fields in stellar cores can be so important that an alternative SN explosion mechanism, called magneto-rotational explosion, was suggested soon after the neutrino mechanism was proposed \citep{1971SvA....14..652B, 1970ApJ...161..541L} (see, e.g., \citealt{2005MNRAS.359..333A} for results of 2D MHD simulations).
Recent 3D MHD numerical simulations \citep[e.g.][]{2015arXiv151200838M} suggest that rapidly
rotating cores of massive stars with magnetic field may trigger the most energetic astrophysical 
explosions including long gamma-ray bursts and super-luminous supernovae. 
Non-axisymmetric instabilities of rapidly rotating proto-NSs may be interesting sources of 
astrophysical gravitational waves (see a summary of recent studies and prospect for future 
advanced LIGO-VIRGO searches in \citealt{2016PhRvD..93d2002G, 2016arXiv160501785A}). 
GWs (gravitational waves) are also promising source of the direct probe of rapid rotation in nucleus of core collapse supernovae
through observation of circular polarization of GW signal from possible 
hydrodynamic instabilities at the post-bounce phase \citep{2016PhRvL.116o1102H}.

In extreme cases, the fast rotation of the iron-oxygen core of a massive star prior to collapse can lead to the formation of dynamically 
unstable proto-NSs \citep{2001ApJ...550L.193C, 2007PhRvD..75d4023B}. For instance, in classical incompressible Maclaurin spheroid approximation \citep{1969efe..book.....C, 1985prpl.conf..534D}, when the ratio of the rotational energy $T$ to gravitational energy $W$, $\beta=T/|W|> \beta_{dyn}$ achieves the critical value $\beta_{dyn}\simeq 0.27$\footnote{The analysis in full GR shows \citep{2007PhRvD..75d4023B} that 
the classical result for incompressible Maclaurin spheroids remains valid, with $\beta_{dyn}\sim0.25-0.27$; here we will use the 
classical value $\beta_{dyn}=0.27$.}, the bar-mode instabilities are formed and can consequently lead to fission of the proto-NS into two parts. As a result, a close binary NS may be formed during the core collapse.
The orbital evolution driven by GW emission brings the lightest (having higher radius) component into contact
with Roche lobe leading to 
the mass transfer from the lightest onto the more massive component. When the mass of the lightest companion decreases down to a minimum possible mass of stable NS, $\sim 0.1\,M_\odot$, density of the matter (and hence Fermi-momentum of electrons) becomes too low 
and enables neutrons in neutron-rich nuclei to beta-decay in a fraction of a millisecond, 
which ends up with an explosion of the $\sim 0.1 M_\odot$ NS remnant 
producing a powerful neutrino outburst.
This scenario was proposed by \cite{1992SvAL...18..194S} to explain unusual neutrino signals from SN 1987A; (see \citealt{1992SvAL...18...79I}, \citealt{2011PhyU...53.1081I} for a review and references). The approximate analytic solution of the
the GW signal (waveform) emitted by these types of the sources was found in \citep{1998AstL...24..206I, 2008AstL...34..375I}. During the first stage of merging, the amplitude and frequency of the GW signal increases, achieving the maximum value in the beginning of the second stage, when 
the low-mass component fills its Roche Lobe.

The idea of rapidly rotating collapsing core fragmentation into two pieces has several important 
astrophysical implications. For example, this can lead to high kick velocities of a NS or a black hole
resulting from the ultimate merging of two NSs \citep{2002ApJ...581.1271C}, 
it may be important in some long gamma-ray burst progenitor models \citep{2009MNRAS.397.1695L} and can be additional source of GW
signals from merging binary NS system \citep{1998AstL...24..206I, 2002ApJ...579L..63D}.  
The latter is especially important for the ongoing searches of GW by the present-day advanced
GW detectors \citep{2015CQGra..32k5012A, 2015CQGra..32b4001A, 2016arXiv160501785A}. 
From the point of view of GW detection, the precise form of the 
GW signal and astrophysical rate of events is of major importance \citep{lrr-2009-2}. While the former 
is very well studied for coalescing binary NSs \citep[e.g.][]{lrr-2014-2}, the event rate of double NS coalescences suffers from many 
astrophysical uncertainties and model assumptions \citep{2003MNRAS.342.1169V, lrr-2014-3,2015ApJ...814...58D}. 

The main motivation of the present study is to calculate, by means of the population synthesis method, the
expected formation rate of rapidly rotating cores of massive stars with parameter $\beta$ 
high enough 
that can potentially lead to noticeable astrophysical 
manifestations. We will focus only on the angular momentum of the core and will not consider 
magnetic fields, which are prerequisite for magneto-rotational and magnetar-powered supernova explosions
and long GRBs. To this end, we will treat the rotation of the stellar core in single and binary massive 
stars using the effective coupling between the core and envelope of the star \citep{1998A&A...333..629A}.
We check the applicability of such a simplified description to the stellar core rotation by comparing
the results with calculations using MESA evolutionary code \citep{2011ApJS..192....3P}. From comparison of the
calculated initial NS rotational periods with the distribution derived from pulsar observations, we 
find the effective coupling constant between the core and envelope (coupling time $\sim 5\times 10^5$~years).
With this coupling, we calculate the distribution of iron core rotation (the $\beta$-parameter) in 
single and binary stars. In single stars, the rotation of the core is too slow for 
strong rotational instabilities to develop. In binary systems, the rapid rotation arises 
at stages of coalescence of non-degenerate stellar cores during the common envelope stage.    
In our calculations we take into account tidal synchronization of the stellar envelope in close binaries
and loss of stellar envelope's angular momentum by stellar wind, as implemented in the BSE population synthesis code
\citep{2000MNRAS.315..543H, 2002MNRAS.329..897H}. 
This is a novel element compared to some previous population synthesis studies, which assumed rigid rotation of the entire star 
in a close binary system \citep[e.g.][]{2009ARep...53..325B, 2009MNRAS.397.1695L}.
We find that for the standard parameters of close binary evolution the fraction of rapidly rotating cores 
can reach 0.1-1\% of the total number of core collapses. This rate is comparable to an optimistic 
Galactic rate of double NS coalescences of about $10^{-4}$ per year \citep{2010CQGra..27q3001A, lrr-2014-3}. 

The structure of the paper is as follows. In Section \ref{s:core_rot} we briefly describe the 
stellar core rotation and its implementation into the modified BSE code.
 In Section \ref{s:initial_NS} we discuss
the initial rotation periods of NSs as derived from pulsar studies. 
In Section \ref{s:results} we presents the results of our calculations, and
Section \ref{s:summary} gives summary of the present study. 
In Appendix A we compare the results of core rotation calculation of a 15 $M_\odot$ star 
with detailed calculation by the MESA code. Appendix B presents the typical evolutionary 
tracks of binary stars that lead to rapidly rotating stellar cores prior to collapse.

\begin{figure}
\includegraphics[width=0.48\textwidth]
{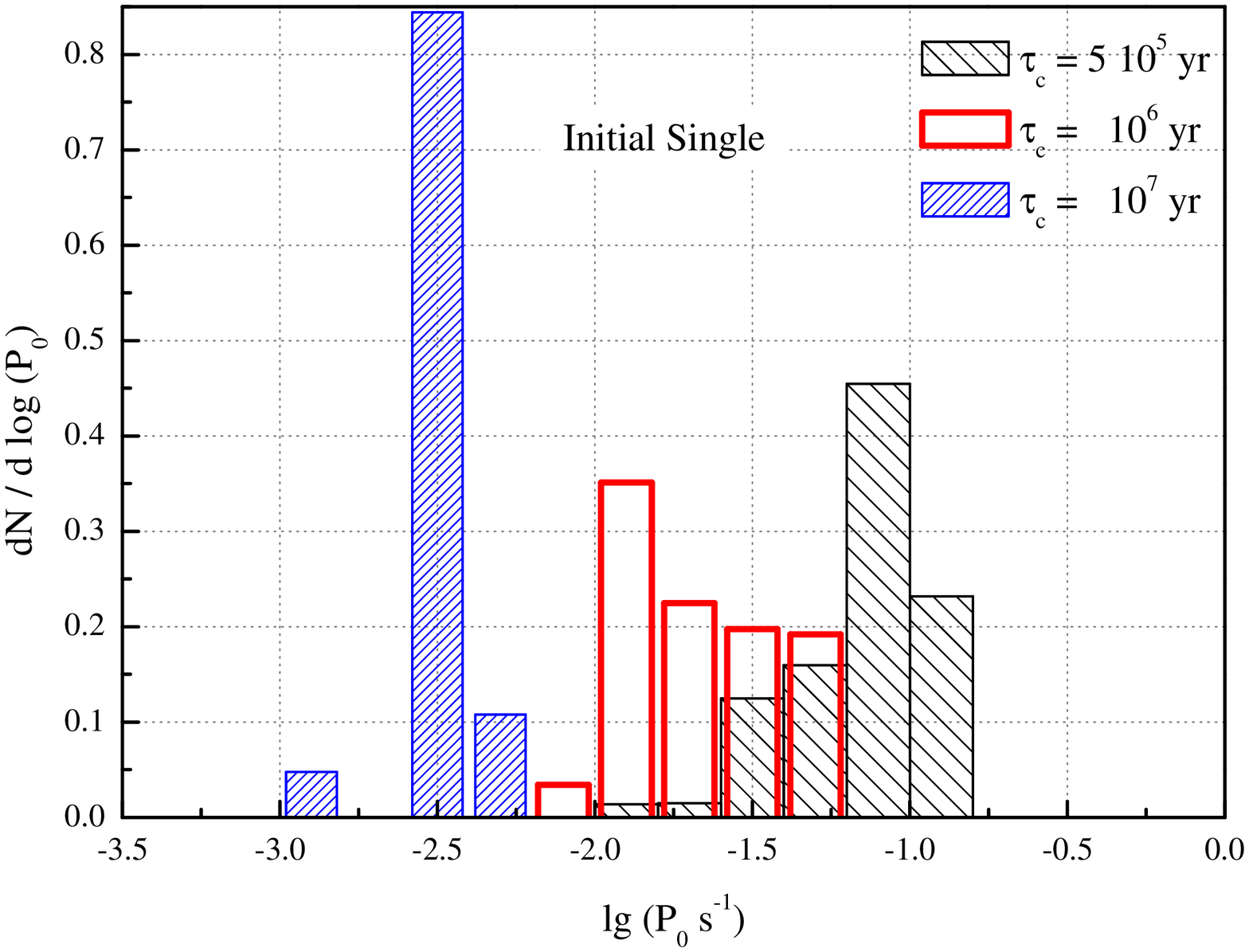}
\vfill
\includegraphics[width=0.48\textwidth]
{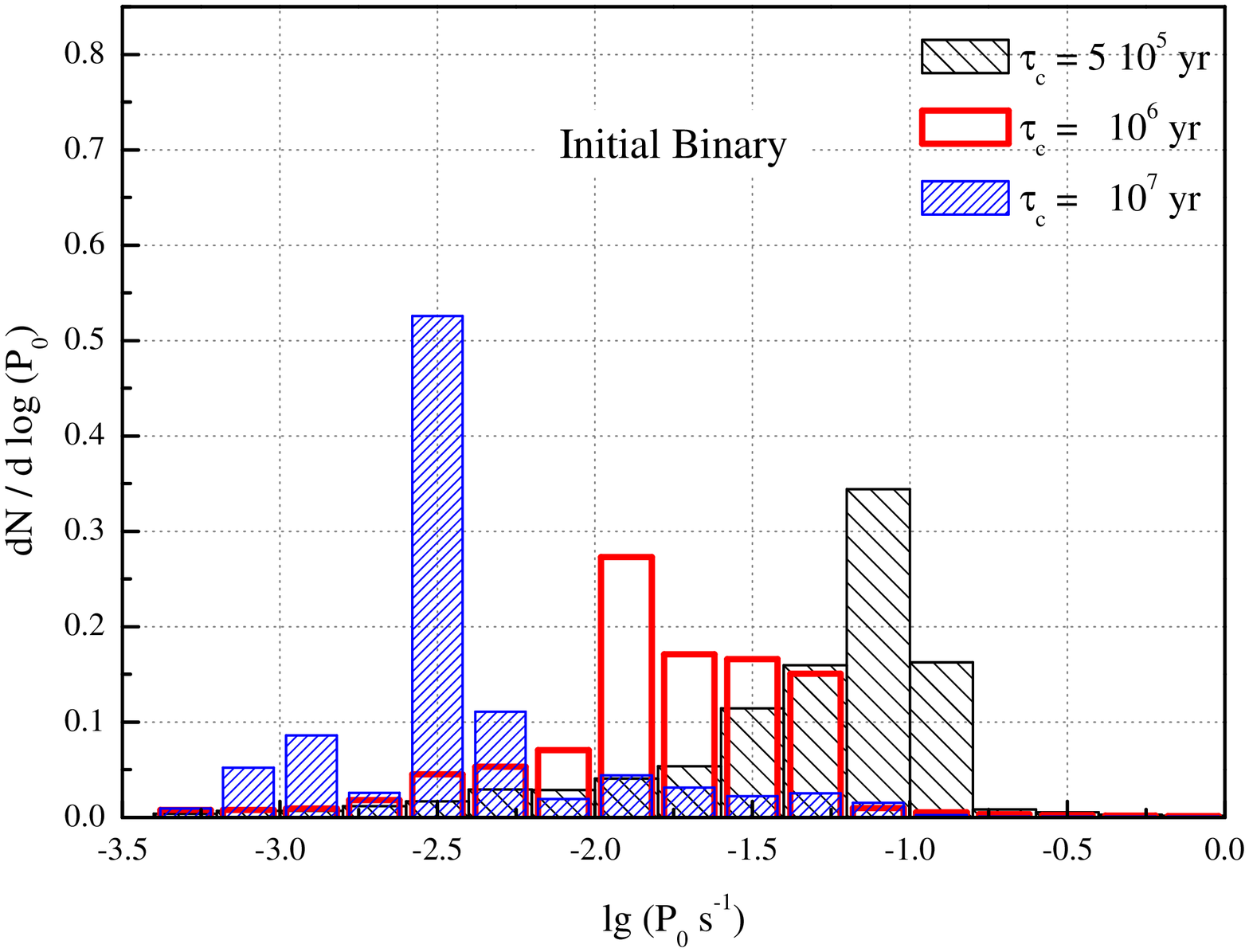}
\vfill
\includegraphics[width=0.48\textwidth]
{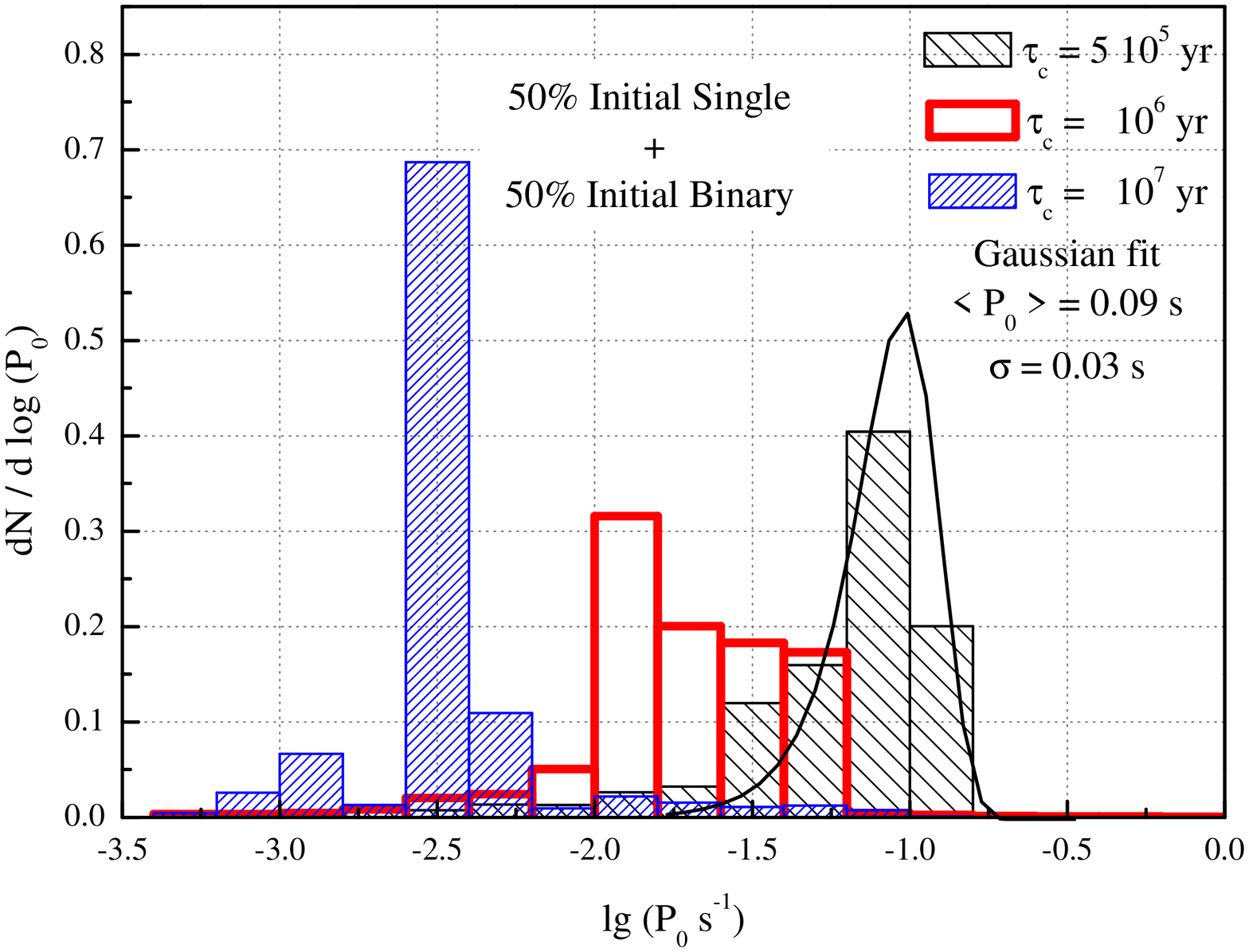}
\caption {Initial spin periods of NSs originated from single, binary and equal number of single and binary progenitors (upper, 
middle and bottom panel, respectively), for different values of the core-envelope coupling time $\tau_c=5\times 10^5$, $10^6$
and $10^7$ years. The solid line in the bottom panel shows the best-fit Gaussian for the calculated $dN/d P_0$ distribution
with parameters shown in the inset. 
\label{p_ns}}
\end{figure}

\section{Description of the stellar core rotation}
\label{s:core_rot}

Stellar rotation is a very broad and topical field of astrophysical studies and is recognized to
be a key factor, in addition to the mass and chemical composition, determining stellar 
evolution \citep{2000ARA&A..38..143M}.
The core-envelope coupling can be 
due to magnetic torques \citep[e.g.][]{2002A&A...381..923S, 2015arXiv151003198B} or internal gravity waves 
excited by convective motions \citep{1997A&A...322..320Z}. 
  
In the present study, the rotational evolution of the stellar core is effectively described as follows. 
Consider a star consisting of core and envelope with moment of inertia $I_c$ and $I_e$, 
respectively. The total angular momentum of the star is $J=I_c\omega_c+I_e\omega_e$, 
where $\omega_c$ and $\omega_e$ are core's and envelope's angular velocity, respectively,
assuming rigid rotation. Following \cite{1998A&A...333..629A}, we assume that the stellar core with moment of inertia $I_c$ is effectively coupled 
with the stellar envelope with moment of inertia $I_e$ (e.g. due to magnetic torques \citep[e.g.][]{2002A&A...381..923S, 2015arXiv151003198B} or internal gravity waves 
excited by convective motions \citep{1997A&A...322..320Z}) as follows:
\beq{e:coupling}
\frac{dJ_c}{dt}=-\frac{I_cI_e}{I_c+I_e}\frac{\omega_c-\omega_e}{\tau_c}
\eeq
where $J_c$ is the angular momentum of the core and  
$\tau_c$ is the core-envelope coupling time. If the latter is too long (e.g., comparable to the characteristic 
time of stellar evolution), the stellar core essentially 
evolves conserving its own angular momentum. 

The initial rotational periods of main-sequence stars were adopted as in the BSE (Binary Star Evolution)
code \citep{2000MNRAS.315..543H}
which uniquely relates the equatorial rotational velocity of a zero-age main-sequence star with its mass, $v_{rot}(M)$ (see
Eq. (107) in that paper).  
Initially, the star is assumed to rotate rigidly, $\omega_c=\omega_e$. During evolution, the core and envelope rotation
changes due to variation of the core and envelope moment of inertia, tidal and stellar wind torques applied to
the envelope as implemented in the original BSE code, and the core-envelope coupling discussed above.  
However, we find that that the initial rotation does not affect the fraction of 
rapidly pre-collapse stellar cores, which is the main purpose of our study.

The comparison of our effective description of the stellar core rotation with
more detailed calculations by MESA code are presented in the Appendix A. 

To quantify the core rotation before the collapse, we will use the ratio of the
rotational energy to the gravitational potential 
energy of the core, $\beta=T/|W|$. The kinetic energy of a rigidly rotating core is  
\begin{equation}
T = \frac{k M_c R_c^2 \omega^2}{2}\,,
\label{eq:T}
\end{equation}
where $M_c$ and $R_c$ is the mass and radius of the core, respectively, and 
$k$ is the numerical factor depending on the density distribution. For example, for polytropic structure with $n=3/2$,
$k\approx 0.21$, and for $n=3$, $k\approx 0.075$. The stellar core
before collapse is be described by a polytropic configuration with index $n=3$, and the 
gravitational potential energy for such a core is 
\begin{equation}
W = -\frac{3}{5-n}\frac{G M_c^2}{R_c}\,,
\label{eq:W}
\end{equation}
where $G$ is the Newton gravitational constant.


\section{Initial neutron star spin periods}
\label{s:initial_NS}

In the present study, we will compare the initial period distributions of NSs  
calculated by the BSE code for different values of the core-envelope coupling parameter $\tau_c$ with 
the initial NS spin period distribution $dN/dP_0$ as inferred from observations. The latter 
requires a separate discussion. 

From the observational point of view, there are several populations of young
isolated NSs: radio pulsars, magnetars (soft gamma-ray
repeaters and anomalous X-ray pulsars), central compact objects (CCOs) in
SNRs, rotating radio transients (RRATs), and so-called X-ray dim isolated NS,
aka the 'Magnificent seven'.  Despite the fact that all of these sources are
relatively young NSs and in many cases rough age estimates can be obtained,
it is very difficult to estimate the initial spin distribution for the
majority of them. There are three main difficulties:
\begin{itemize}
\item It is necessary to have a precise model of the NS spin-down;
\item It is necessary to know how several key parameters evolve 
(magnetic field and magnetic inclination angle);
\item It is necessary to have precise (better than a few 10\%)
independent age estimate.
\end{itemize}

Significant evolution of the magnetic field can be expected in magnetars. Therefore
in this case the determination of $P_0$ is the
most difficult task. Similar conclusions can be made about the Magnificent
seven, as they also represent evolved NSs with initially large magnetic
field (\citealt{popov2010}). 
Thus, CCOs and radio pulsars (RRATs can be considered together with radio pulsars) seem to
be the most appropriate objects for estimates of the initial NS spin periods. 

In a few cases where NSs are related to historical SNe, 
the NS age is known rather accurately \citep[e.g.][]{1982ASIC...90..355C,2000PhyU...43..509I}. 
However, other estimates are required to get a larger statistics.
There are two main approaches to obtain independent age estimates for tens of NSs.
The less precise one is from kinematic considerations. The so-called kinematic
age have been used by \cite{noutsos2013} to derive $P_0$ for several tens of radio pulsars. 
The second, more precise approach is based on the NS age estimation from the SNR observations. This method 
has been used by many authors. For example, in the recent attempt \cite{pt2012} about 30 NS+SNR pairs were analyzed.

In addition, population synthesis modeling have also been used to probe the 
initial NS spin period distribution \citep[see, e.g.][]{2006ApJ...643..332F, popov2010, gullon2014,
gullon2015}. 

In brief, all studies (using samples of sources with known ages and
population studies) demonstrate that radio pulsars (which dominate the 
samples) can have relatively
broad initial spin period distribution up to 0.5-1 second, and typical
periods are about 100 ms. 

\cite{pt2012} did not try to produce an exact distribution, as the available 
statistics of objects is
not large enough. But it was possible to probe different popular analytical
distribution vs. cumulative initial spin period distribution derived from
PSR+SNR pairs. The authors 
demonstrated that both very broad flat and very narrow ($\delta$-function like) distributions do not fit well the
data. On the other hand, Gaussian distributions with $\langle P_0
\rangle\sim 0.1$-0.2~s and $\sigma\sim 0.1$-0.2~s are found to be in agreement 
with the data. 

\cite{noutsos2013} produced the initial spin period distribution using
several tens of radio pulsars with known kinematic ages. Their distribution
is bimodal. One mode is in correspondence with findings by \cite{pt2012}. In
addition, they proposed that there is a ``tail'' of young NSs with initial
spin periods $P_0\sim 0.5$-0.7~s.
However, the latter feature can be a result of some evolutionary effects (magnetic field
decay or emergence) which have not been taken into account in that study \citep[see][]{ip2013}.

In population synthesis studies, different authors infer 
the NS initial spin period distribution from fitting the present-day radio pulsar
properties by model distributions. This approach is based on many assumptions, and it is
difficult to derive an independent exact form and parameters of the $P_0$
distribution. For example, the possibility of ``injection'' of
pulsars at spin periods $\sim$0.5~s cannot be rejected (\citealt{1981JApA....2..315V, vran2004}). 
Generally, results of different studies are not in contradiction with analysis of samples of young
NSs with known ages. Initial spin periods $\sim100$~ms are favored, but
the distributions are not very narrow, and allow some pulsars to be born with ms
periods, and some --- with $P_0\sim$0.5~s.  

Theoretical estimates of the $P_0$ distribution remain rather uncertain (see recent
reviews and references in \citealt{miller2015, fuller2015}). Several 
mechanisms can significantly affect the initial NS rotation. In
particular, the standing accretion shock instability (SASI) can modify the
NS spin (\citealt{sasi2014} and references therein). Effect of fall-back accretion  onto 
the nascent NS can also be important \citep{2001ApJ...554L..63M}.

Therefore, in the present study we will assume, guided by observations, that 
initial NSs periods have a Gaussian-like distribution centered on $\sim 100$ ms
with a comparable dispersion.

\section{Results}
\label{s:results}

\subsection{Core-envelope coupling time}
First, we performed BSE population synthesis calculations of the initial NS spin periods
from single and binary progenitors. 
A NS was assumed to be formed in the ZAMS mass range from $M_{min}=8 M_\odot$
to $M_{max}=35 M_\odot$.  
The rotation of the stellar He and CO-cores was treated as described above. The mass of the pre-collapse iron-oxygen core
was assumed to be 1.8~$M_\odot$, and its rotation was calculated by assuming angular momentum
conservation after the CO-core stage. In other terms, a mass of 1.8~$M_\odot$ was cut from the 
CO-core described by $n=2$ polytrope (which is a reasonable approximation at the beginning of the CO-stage) 
with the angular momentum $J_{1.8}=I_{1.8}\omega_{CO}$, 
where $I_{1.8}$ is the moment of inertia of the central 1.8 $M_\odot$ part of the CO-core and $\omega_{CO}$ is
the angular velocity of the CO-core at this stage. Then the rotation of the
iron-oxygen core was calculated assuming the 1.8 $M_\odot$ $n=2$ rigidly rotating polytrope
contracting to $n=3$ rigidly rotating polytrope (which is good approximation of the 
core sptructure prior to collapse) with the same mass and angular momentum, i.e. 
$\omega_{Fe}=\omega_{CO}I_{1.8}(n=2)/I_{1.8}(n=3)$.   

During the core collapse, the angular momentum conservation
was assumed, $I_{1.8}\omega_{Fe}=I_{NS}\omega_{NS}$, where the NS moment of inertia $I_{NS}=10^{45}$~g~cm$^2$. 
Therefore, the initial NS period was found as $P_0=(2\pi/\omega_{NS})$. 
(At this stage we neglect possible dynamical instabilities of the rapidly rotating stellar cores and young NSs.) 
The binary progenitors include all binary systems in which the initial 
primary's mass $M_{max}\ge M_1\ge M_{min}$. 
The standard binary population synthesis initial parameters were adopted: a flat 
distribution of the binary mass ratio $q=M_2/M_1\le 1$: $dN/dq\sim q^0$; a flat distribution of the 
initial binary separation, $dN/d\log a=$const and the distribution of initial binary periods
by massive binaries proposed by \cite{2012Sci...337..444S}, $dN/(d\log P)^\pi=$const with $\pi=-0.55$; 
alpha-description of the common envelope (CE) stage with 
the efficiency parameter $\alpha_{CE}$ (see, e.g. \citealt{lrr-2014-3} for a detailed description of binary evolution 
and population synthesis parameters). Note that the natal kick velocity imparted to NS in core collapse is
not important for our calculations as long as central kicks are considered that do not affect the NS spin
(see, however, \citealt{1998Natur.393..139S} for alternative explanation of NS spins, which we do not consider here).

The resulted (normalized) distribution of NS initial periods
$P_0$ is shown in Fig. \ref{p_ns} for a population of single, binary and equal fractions of single and binary progenitors 
(the upper, middle
and bottom panel of Fig. \ref{p_ns}, respectively) for different values of the core-envelope coupling time $\tau_c=5\times 10^5$,
$10^6$ and $10^7$ years. 
For each set of parameters, evolution of a population of $5\times 10^5$ systems was calculated. 

It is seen that the long (comparable to the stellar evolution) core-envelope 
coupling time essentially leads to 
the core evolution with angular momentum conservation and overproduction of rapidly rotating NSs. The evolution 
in binary systems (mostly via tidal interaction in close binaries) leads to wider $P_0$ distributions, but still essentially 
peaked at the same values as from single progenitors (the middle panel of Fig. \ref{p_ns}), and therefore the 
initial $P_0$ distribution from a population of equal number of single and binary progenitors 
(the bottom panel of Fig. \ref{p_ns}) does not significantly differ from that resulted from binary progenitors. 
In the bottom panel of Fig.  \ref{p_ns}, the calculated $dN/d P_0$ distribution for $\tau_c=5\times 10^5$ years  
can be fitted by a Gaussian
with $\langle P_0\rangle $=0.09~s and $\sigma =0.03$~s (the solid line), which is reasonably close to the distribution 
derived from observations (see Section \ref{s:initial_NS}).
Thus we conclude that the core-envelope coupling time should be close to $5\times 10^5$ years to reproduce 
the initial NS period distribution.  
Note that further decrease in $\tau_c$ would shift the $P_0$ peak to longer periods, in contradiction with
observations. Physically, a shorter $\tau_c$ means more effective coupling between the core and envelope, and the expansion of the envelope during the stellar evolution strongly spins down the stellar core. 
Below we shall take $\tau_c=5\times 10^5$~years. 

\subsection{On the magnetar formation from rapidly rotating cores}

Rapidly rotating NS have been frequently invoked as possible progenitors of
magnetars, --- i.e. highly magnetized NSs, --- 
due to an effective dynamo mechanism, enhancing the magnetic field in a proto-NS 
(\citealt{1993ApJ...408..194T}; see
\citealt{2015RPPh...78k6901T} for a recent review).   It is unclear, however, if indeed a dynamo
mechanism is responsible for generation of inferred high magnetic fields of magnetars, and if yes, which
kind of dynamo is in operation. Therefore, it is difficult to
specify exact initial conditions for a NS to become a magnetar, in particular, 
what should be the critical spin period of the proto-NS. For example, \cite{1993ApJ...408..194T} advocated
the range $\lesssim 3$~ms, but this value can vary in different
dynamo models. 

Spin-up of the stellar core in magnetar progenitors due to binary evolution
(accretion of angular momentum, tidal synchronization, coalescence) have
been discussed by \cite{2006MNRAS.367..732P}. 
There is observational evidence that magnetars SGR 1900+14 and CXOU
J1647-45 may have been born in binary systems.
The first of them was studied by \cite{2009ApJ...707..844D}.
Analysis of the cluster in which SGR 1900+14 was born brought the
authors to the conclusion that the source progenitor was likely a
member of a binary, since otherwise it would be difficult to bring in
correspondence relative low progenitor mass ($17\pm1 \, M_\odot$) with
standard scenarios of stellar evolution and magnetar formation.
The progenitor of the second source was studied by
\cite{2006ApJ...636L..41M} and recently in more details by
\cite{2014A&A...565A..90C}.
In the last paper, the authors provided direct arguments in favour of
the magnetar formation in a binary system and 
concluded that 'binarity is a key ingredient in the formation of at least a
subset of magnetars'.

The fraction of magnetars among newborn NSs is estimated to be $\sim10$\% (see,
for example, \citealt{2015RPPh...78k6901T} and references therein).
The initial spin distributions of NS presented in Fig.~\ref{p_ns} suggest
that for the effective core-envelope coupling time $\tau_c <10^6$~yrs 
the fraction of rapidly rotating NSs (with 
initial spin periods $P_0\lesssim
10$~ms) is much smaller than 10\%, while $\tau_c >10^6$~yrs is in
contradiction with isolated NS data. Therefore, we conclude that
in the framework of our model description of the rotation of stellar cores  
the whole magnetar population cannot originate only from rapidly rotating proto-NSs.

\subsection{Stellar core rotation before the collapse}
\begin{figure}
\centering
\includegraphics[width=0.49\textwidth]
{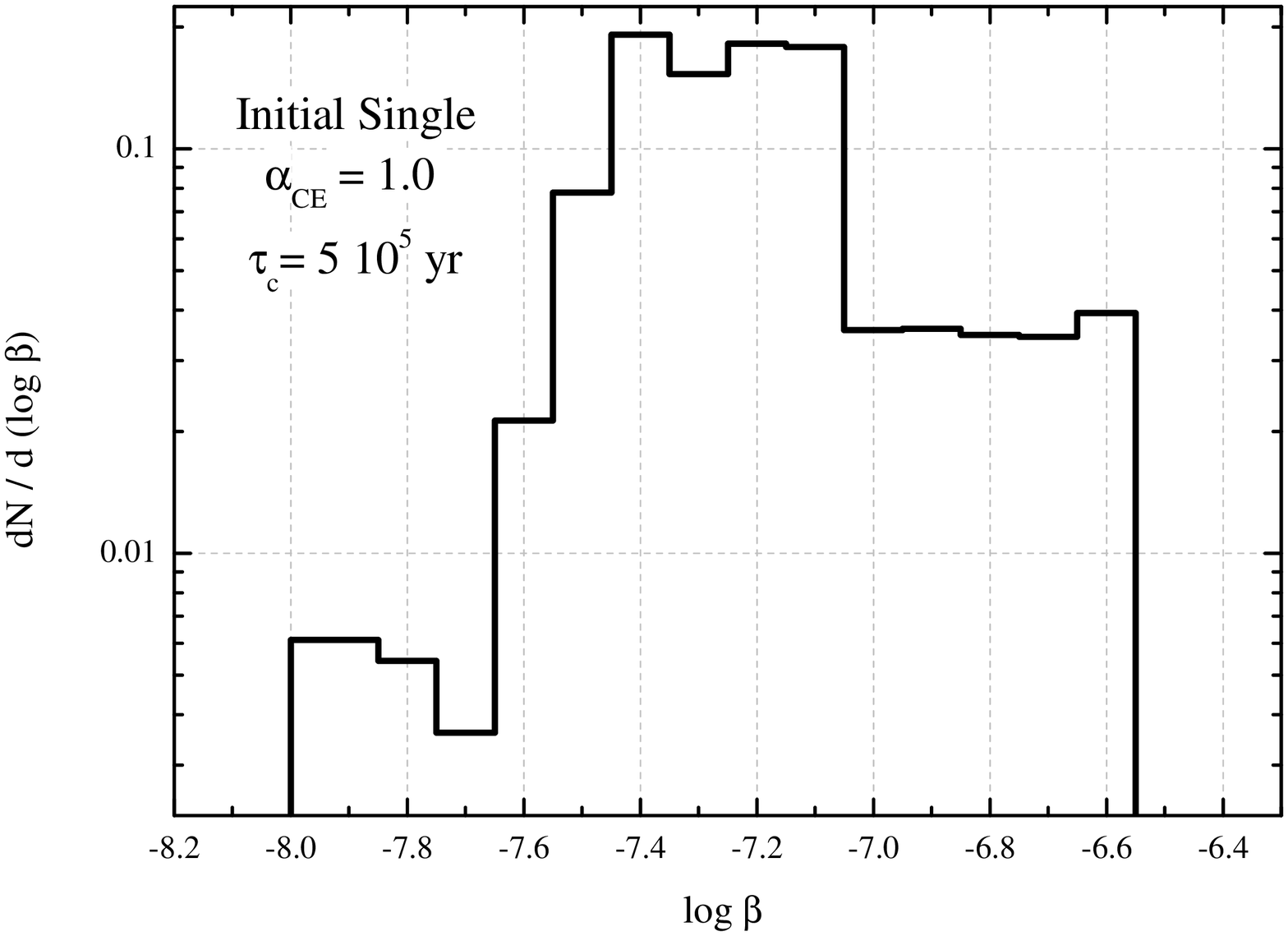}
\vfill
\includegraphics[width=0.49\textwidth]
{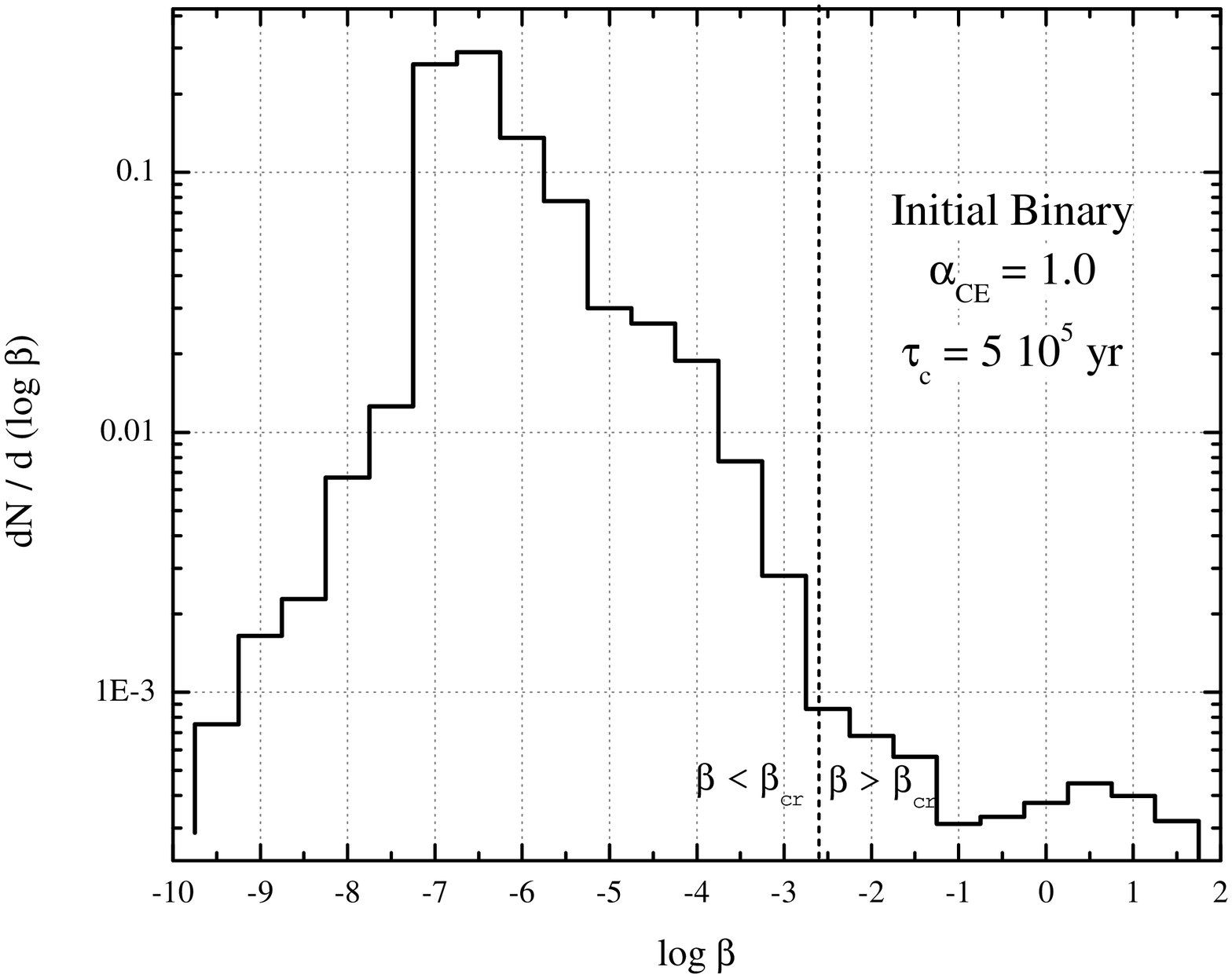}
\caption {
Distribution of the pre-collapse 
core rotation parameter $\beta = T/|W|$ for single and binary NS progenitors.
\label{beta}}
\end{figure}


\begin{figure*}
\includegraphics[width=0.45\textwidth]{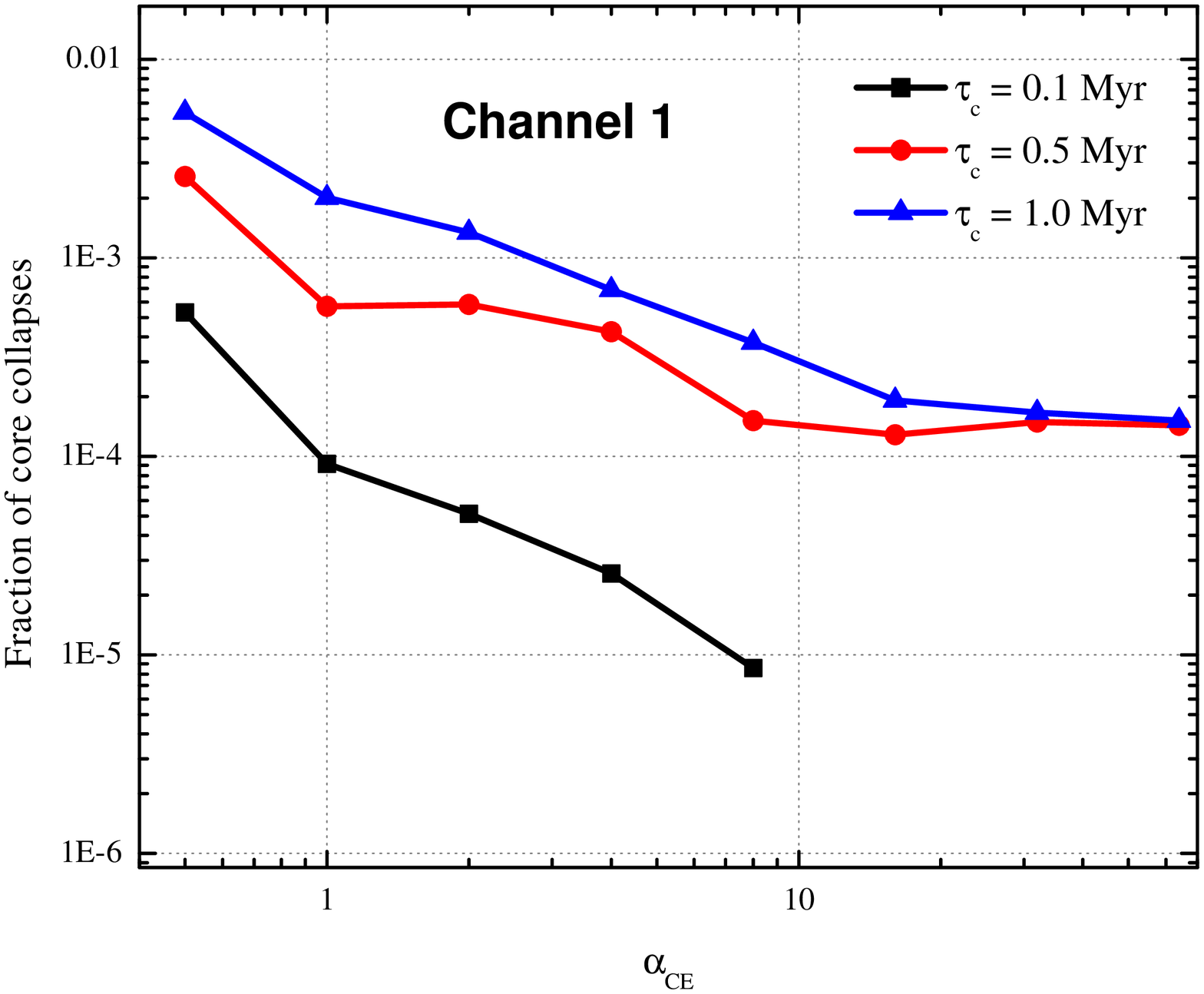}
\hfill
\includegraphics[width=0.45\textwidth]{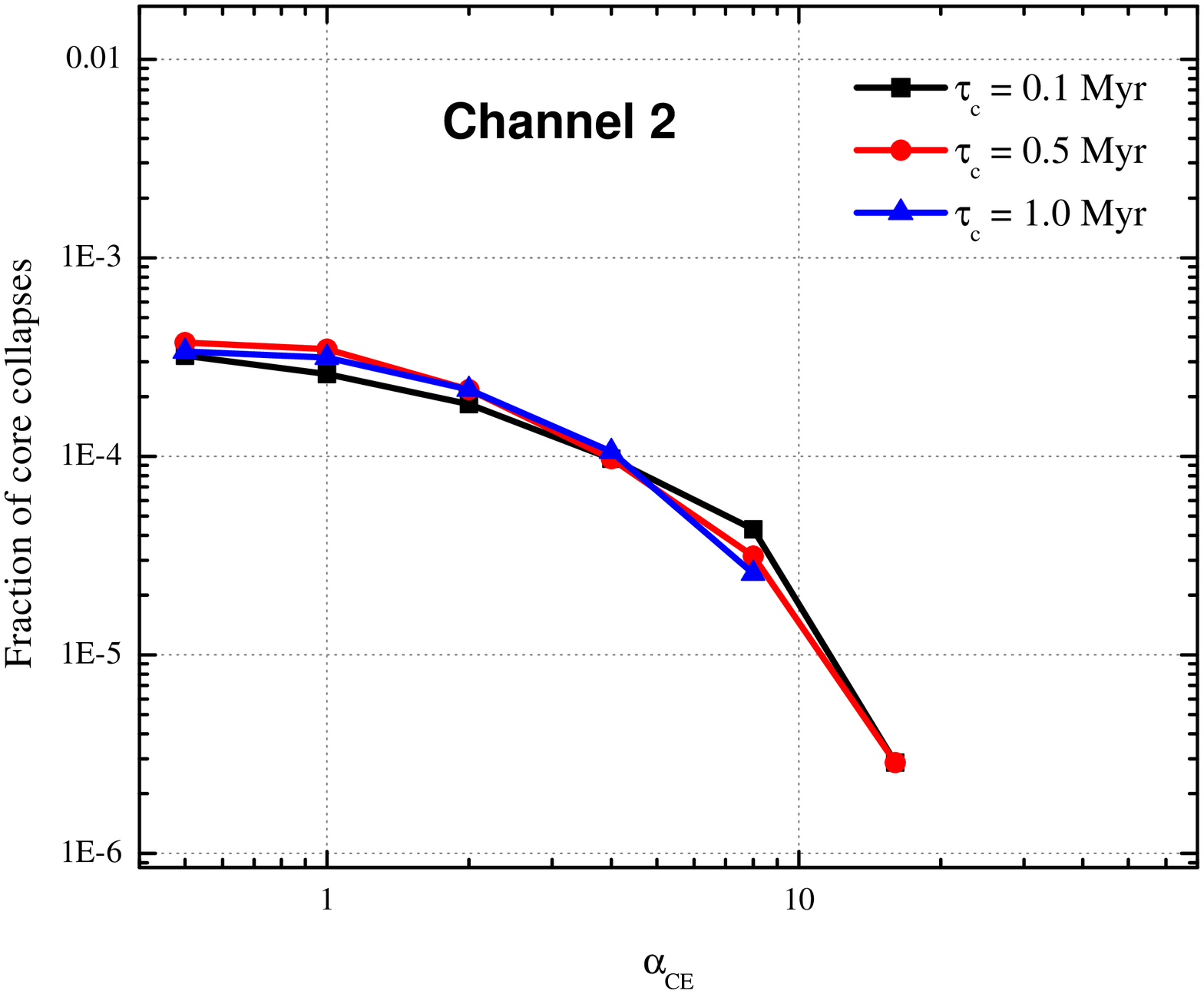}
\vfill
\includegraphics[width=0.45\textwidth]{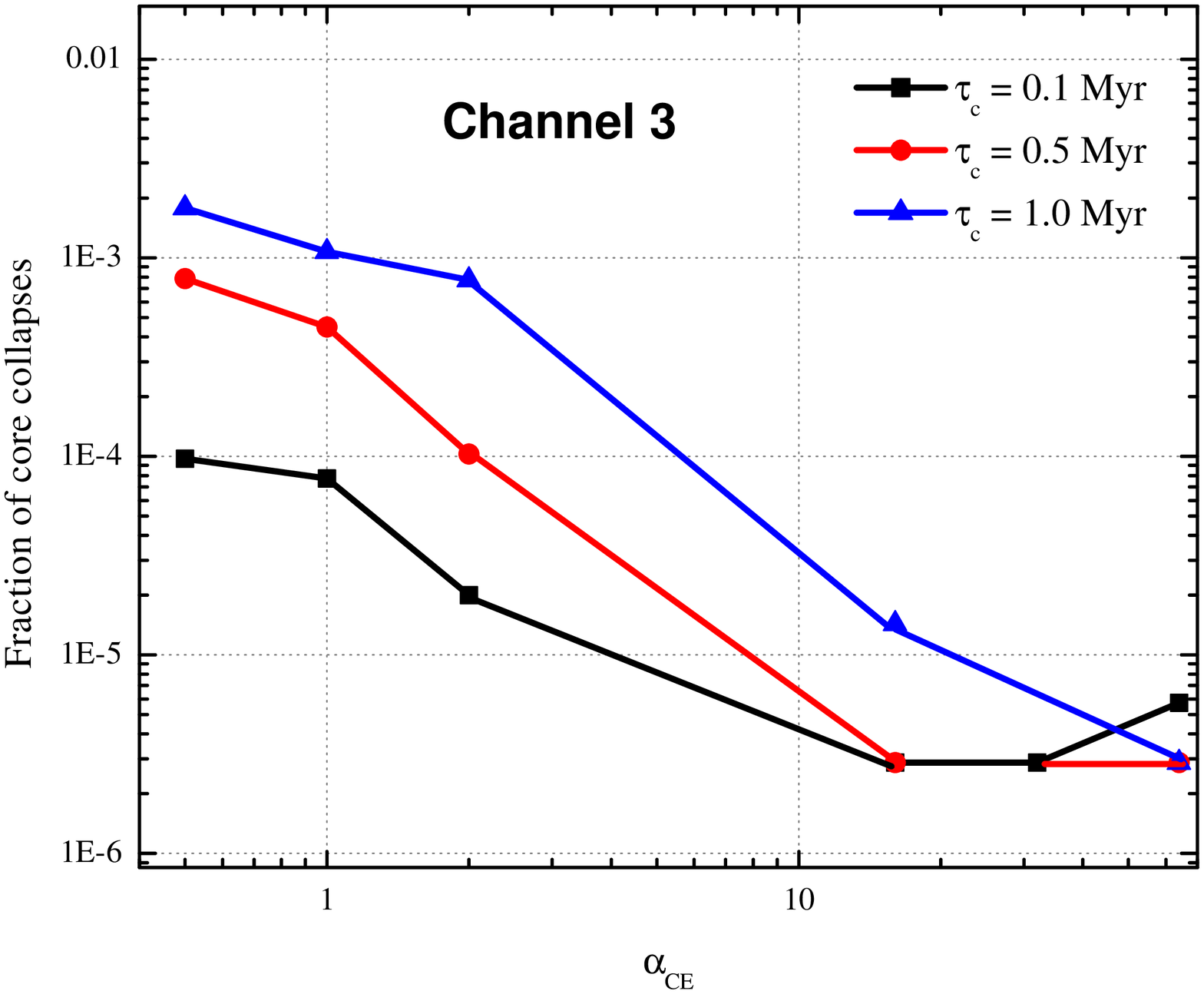}
\hfill
\includegraphics[width=0.45\textwidth]{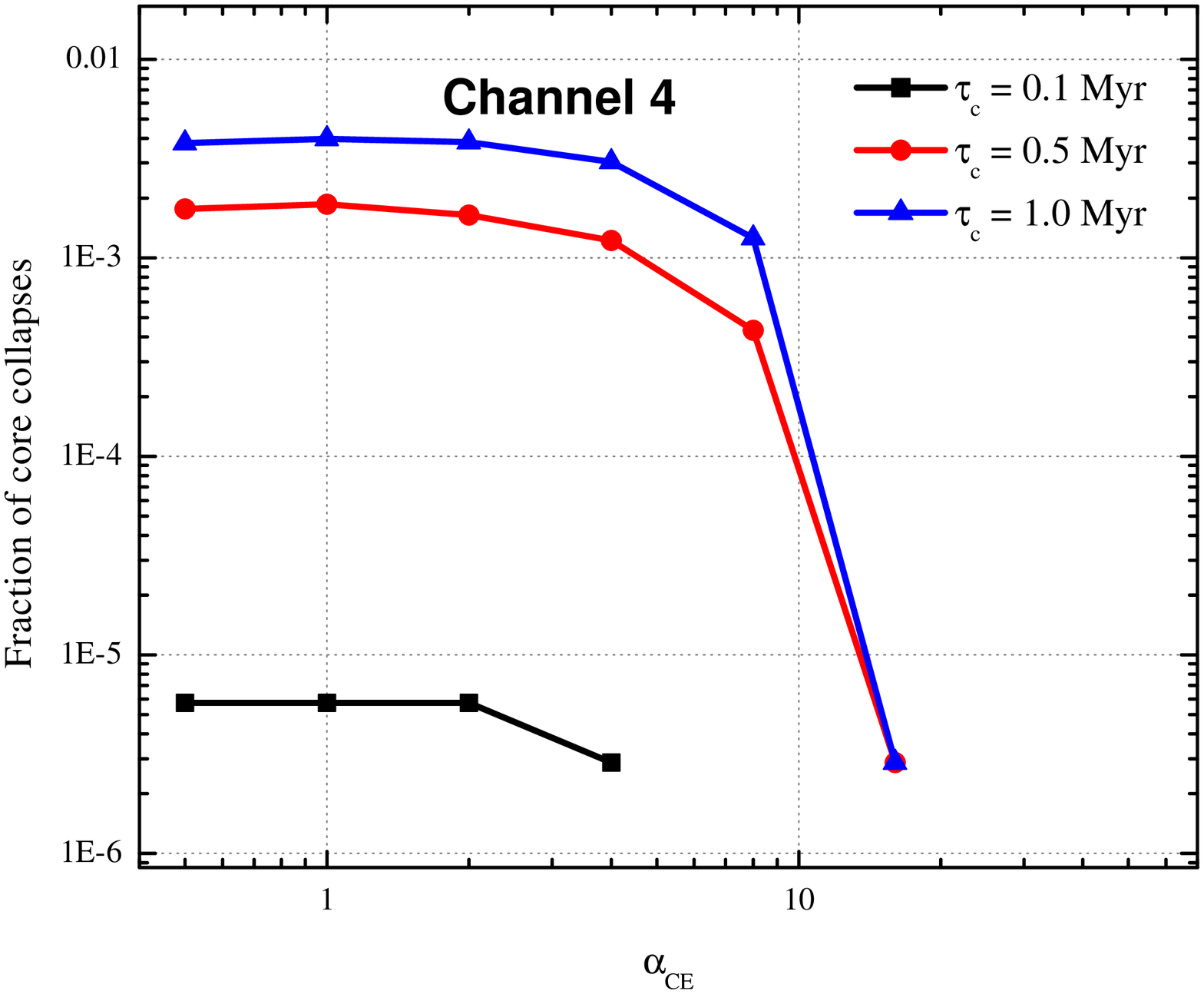}
\vfill
\includegraphics[width=0.45\textwidth]{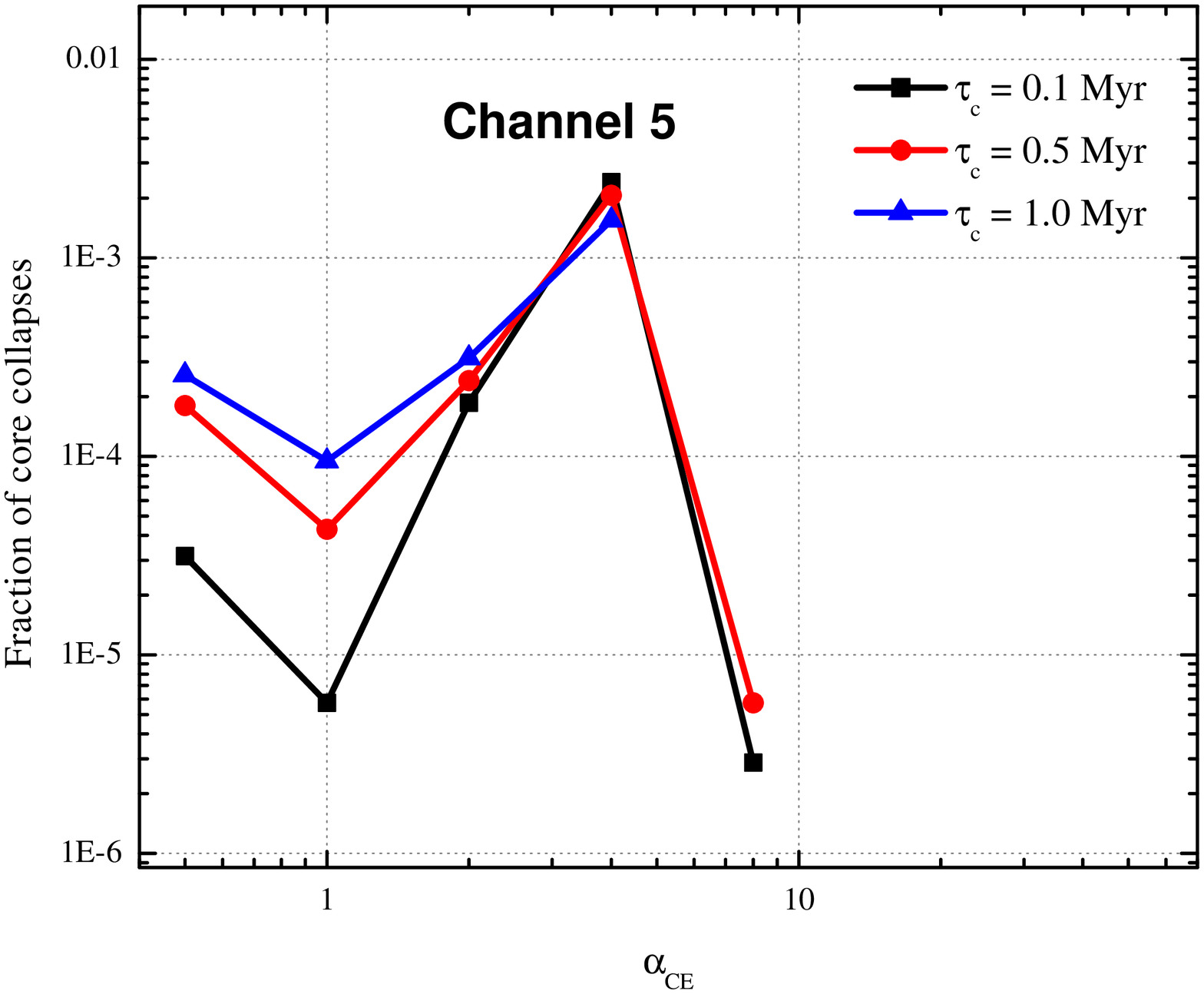}
\hfill
\includegraphics[width=0.45\textwidth]{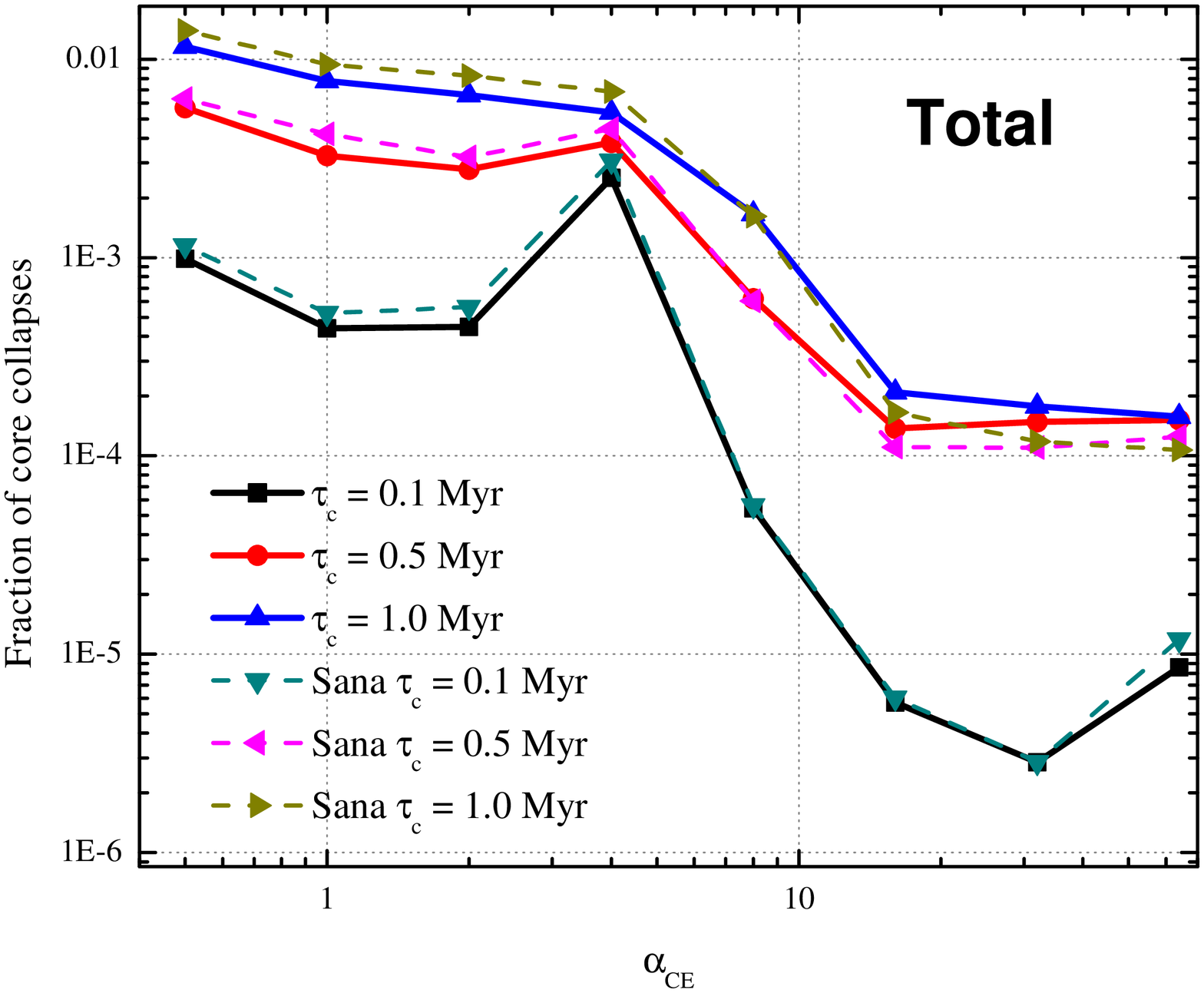}
\caption{Fraction of rapidly rotating cores as a function of the common envelope parameter $\alpha_{CE}$ in different 
formation channels (Tables \ref{tab:track1}-\ref{tab:track5}) for three values of the core-envelope effective coupling
time $\tau_c=0.1, 0.5, 1$~Myr. Bottom right panel shows the sum from all channels leading to core collapses
for two initial binary orbital period distribution -- the flat distribution (solid lines) and Sana's distribution 
\citep{2012Sci...337..444S}. 
\label{f:channels}}
\end{figure*}

In Fig. \ref{beta} we show the distribution of the rotation parameter $\beta$ of the core before the collapse for single 
NS progenitors (the upper panel) and for binary progenitors (the bottom panel) with $\tau_c=5\times 10^5$~years. 
It is seen that the value of $\beta$
for single stars is always too small to allow the dynamical instability to develop. However, for binary progenitors 
the situation is much more interesting. Indeed, we see that there is a tail of rapidly rotating cores 
with high $\beta$. For the dynamical instability to develop, proto-NS should have $\beta>\beta_{dyn}\sim 0.27$. Since during the collapse with angular momentum conservation (by assumption) $\beta$ increases proportionally to the inverse radius of the 
contracting configuration,     
potentially interesting for the dynamical instability will be the cores with $\beta>\beta_{cr}\sim 2.7\times 10^{-3}$, 
where we have conservatively assumed that the radius of a proto NS is 100 times smaller than that of the pre-collapse iron core\footnote{For example, in detailed calculations by \cite{2009AstL...35..799I} the collapsing 1.8 $M_\odot$ rapidly rotating iron-oxygen core described by 
$n=3$ polytrope had a radius of $3.75\times 10^8$~cm.}. 
The bottom panel of Fig. \ref{beta} shows that there can be a non-negligible fraction
of such rapidly rotating stellar cores for the assumed $\tau_c=5\times 10^5$~years. It is interesting to 
see what are the progenitor binary systems of such rapidly rotating cores.

From the present population synthesis calculations we find that all such rapidly rotating cores 
resulted from merging in the common envelope stage. Typical evolutionary tracks 
leading to such merging are listed in Tables \ref{tab:track1}-\ref{tab:track5} in Appendix B. 
The eclipsing binary system GU Mon with early type companions of equal masses which are in contact 
can be an example of the binary system at the stage of the beginning of the common envelope stage \citep{2016arXiv160303177L}.  
If merging of two stellar cores occurs inside the common envelope, 
the angular momentum of the coalesced core, $J_c'$,  is set to be equal to the orbital angular momentum
of the two contacting cores prior to the coalescence, $J_c'=M_{c,1}M_{c,2}\sqrt{G(R_{c,1}+R_{c,2})/(M_{c,1}+M_{c,2})}$, where 
$M_{c,1}, M_{c,2}$ and $R_{c,1}, R_{c,2}$ are masses and radii of the cores, respectively. The rigid initial rotation 
and no coupling with the ejecting envelope is assumed.Most binary models that lead to
the formation of rapidly rotating pre-SN cores start with mass ratio $q\sim 1$ (see Tables \ref{tab:track1}-\ref{tab:track5} in Appendix B).
Further rotational evolution of the newly formed helium core is treated in the same way as for ZAMS stars described above
(i.e. the core-envelope structure with effective coupling $\tau_c$ is assumed). 

The fraction of the rapidly rotating pre-collapse iron cores with $\beta>\beta_{cr}\sim 2.7 \times 10^{-3}$ among all core collapses 
are shown in Fig. \ref{f:channels} as a function of the CE efficiency parameter $\alpha_{CE}$ for three effective core-envelope coupling times 
$\tau_c=10^5, 5\times 10^5$~years and $10^6$ years and for different formation channels listed in Tables \ref{tab:track1}-\ref{tab:track5}.
The second parameter describing the common envelope stage in the BSE code, the $\lambda$-parameter, which 
is the measure of the gravitational binding energy of the envelope, was calculated
in all cases according to the BSE code prescriptions \citep{2000MNRAS.315..543H}.
It is seen that depending on the parameters, 
in particular evolutionary channel this fraction varies by about two order of
magnitude from a few $\times 10^{-5}$ to a few $\times 10^{-3}$, with larger fraction for more effective CE (i.e., with smaller $\alpha_{CE}$-parameter). The non-monotonic dependence of the formation rate in channels 3 and  5 on the CE efficiency parameter is due to the occurrence of several CE stages.   
Note that not all formation channels are possible at particular values of $\alpha_{CE}$. It is also seen that generally, with increasing 
$\tau_c$ the fraction of rapidly rotating cores increases as expected. Due to
the weakening of the core-envelope angular momentum coupling, the envelope, which mainly expands and looses the momentum in stellar winds, does not effectively brake the stellar core.
The total formation rate of rapidly rotating cores from all evolutionary channels (see the bottom right panel in Fig. \ref{f:channels}) can 
be $10^{-3}-10^{-2}$ of all core collapses. The assumed initial mass ratio distribution of the binary components does not alter this result.

\section{Summary and conclusions \label{s:summary}}
\label{s:summary}

In the present study, using the population synthesis method based on the BSE code, we calculated the expected fraction 
of the rapidly rotating pre-collapse stellar cores, which potentially can lead to the formation of 
dynamically unstable proto NSs. The critical rotation of the pre-collapse cores is quantified by
the ratio of the rotational to gravitational energy, $\beta=T/|W|$. Potentially dynamically unstable proto-NS 
arises from cores with $\beta>\beta_{cr}\sim 2.7\times 10^{-3}$. The collapse of such a rapidly rotating cores can 
result in fission of the core in two parts with the formation of binary proto-NS. Coalescences of such close binary NSs 
can be potential gravitational wave sources. To calculate the formation rate of such rapidly rotating stellar cores, 
the treatment of the core rotation in two-zone approximation was used. In this approach, the stellar core is assumed to be
coupled with the envelope by \Eq{e:coupling}, which has one free parameter -- the effective coupling time 
$\tau_c$. The applicability of this simplified treatment of stellar core rotation was checked by 
direct calculation of evolution of a rotating 15~$M_\odot$ star by MESA code (Appendix A).

To fix the core-envelope coupling time, we have calculated the expected distribution of initial spin periods 
of NSs formed from single and binary star evolution and compared it with the spin distribution of young NSs
as inferred from pulsar statistical studies \citep{pt2012}. We have found that $\tau_c=5\times 10^5$ years gives reasonable initial 
NS spin period distribution centered at $\sim 0.1$~s. Then the BSE population synthesis code supplemented with the core-envelope
rotational coupling treatment by \Eq{e:coupling} was used to calculate the expected distribution of 
the core rotation parameter $\beta$ prior to the collapse (Fig. \ref{beta}). It was found that rapidly rotating core 
with $\beta>2.7\times 10^{-3}$ may arise from binary evolution. The tail of rapidly rotating cores appears as a result of 
mergings of two non-degenerate 
stellar helium cores during common envelope stages in several evolutionary channels (Tables \ref{tab:track1}-\ref{tab:track5}
in Appendix B). The angular momentum of the merged core is mostly determined by the orbital motion of
the stellar cores in the common envelope, and hence is insensitive to the assumed initial distribution of
stellar rotation at the ZAMS stage.  
At a fixed $\tau_c$, the common envelope efficiency parameter $\alpha_{CE}$ mostly affects 
the formation rate of rapidly rotating cores. We have found that depending on $\alpha_{CE}$, the fraction of 
rapidly rotating cores among all core collapses for all possible evolutionary channels can reach 0.1-1\%,
which can be interesting from the point of view of the formation of double proto-NS from core collapses
as possible gravitational wave sources. 

Clearly, our calculations cannot definitely predict the formation rate of binary proto-NS, and should be considered 
only as upper limits. Future 3D calculations of the collapse of rapidly rotating stellar cores (K. Manukovsky et al., 
in progress) should clarify the fate of the collapse and the possibility of the formation of binary proto-NS through this 
evolutionary channel.  
     
\section{Acknowledgements}

The authors acknowledge S.I. Blinnikov and K.V. Manukovsky for discussions and Dr. Thomas Tauris for
critical remarks. 
The work of KP and NP (study of stellar rotation, writing the paper, analysis of the results) is supported by RSF grant 14-12-00203. 
The work of AK, DK and SP (numerical simulations, checking with MESA code, analysis of the initial NS spin periods) 
is supported by RFBR grant 14-02-00567. The authors acknowledge support from M.V. Lomonosov Moscow State University Program of Development.

\begin{figure}
\centering
\includegraphics[width=0.49\textwidth]
{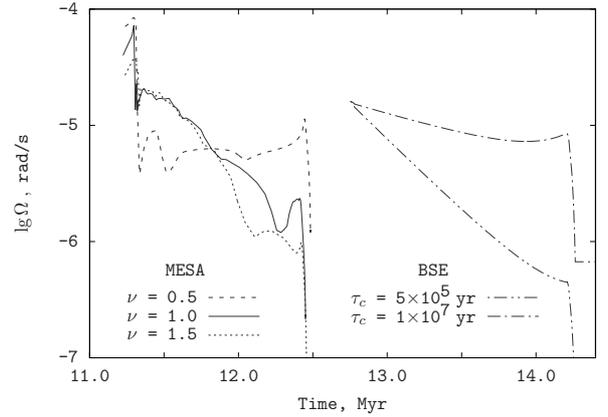}
\caption {The comparison of the time evolution of the angular rotational velocity of the helium core of a 15 $M_\odot$ star with initial rigid rotation calculated  by MESA code (left curves) and BSE (right curves)  
for different values of the core-envelope coupling time $\tau_c$.  
\label{f:MESA}}
\end{figure}

\appendix
\section{Comparison with MESA code}
\label{s:appendix}

We have compared our prescription of the rotational core evolution of normal stars (Section \ref{s:core_rot})
with calculations carried out by MESA code \citep{2015ApJS..220...15P}. As an example, we have considered evolution 
of a 15~$M_\odot$ star with solar metallicity and initial rotation velocity given by Eq. (107) of \cite{2000MNRAS.315..543H}.

In the MESA code, 
implementation of rotation closely follows papers \cite{1989ApJ...338..424P, 2000ApJ...528..368H, 2005ApJ...626..350H}
where the angular momentum transport formulates as a diffusive process. There are three adjustable efficiency factors to calibrate diffusion coefficients. The first parameter is the ratio of the turbulent viscosity to the diffusion coefficient, $f_c = D/\nu$. It changes the time scale of the chemical composition transport. The second parameter is $f_{\mu}$ which describes the sensitivity of the rotationally induced mixing to chemical composition $\mu$-gradients. The last parameter is the diffusion coefficient of the angular momentum $f_{\omega}$ which changes the time scale of the angular momentum transport. Following \cite{2000ApJ...528..368H}, 
we set $f_c = 1/30$ and $f_{\mu} = 0.05$. Influence of the parameters $f_{\omega}$ on the models is not significant in our case and we set it equal to unity.

In the MESA code there are several other parameters related to different mechanisms 
responsible for the angular momentum transport across the stellar radius. In Table \ref{t:MESA} 
we list the parameters and their values we have used in our calculations (\verb+http://mesa.sourceforge.net/+).
\begin{table}
\caption{Parameters in the MESA code used in model calculations of He core rotation of a 15 $M_\odot$ star.}
{\scriptsize
\begin{tabular}{lcc}
\hline
\hline
MESA parameter & Value & Comment\\
\hline
\verb+D_DSI_factor+ & 1.0       &          dynamical shear instability\\
\verb+D_SH_factor+  & 1.0       &          Solberg-Hoiland parameter \\
\verb+D_SSI_factor+ & 1.0       &          secular shear instability \\
\verb+D_ES_factor+  & 1.0       &          Eddington-Sweet circulation \\
\verb+D_GSF_factor+ & 1.0       &          Goldreich-Schubert-Fricke parameter \\
\verb+D_ST_factor+  & 1.0       &          Spruit-Tayler dynamo \\
&\\
\verb+am_nu_visc_factor+ & 0.5, 1.0, 1.5 & kinematic shear viscosity $\nu$\\
&\\
\verb+am_D_mix_factor+ &  1/30  &           $f_c$ \\
\verb+am_nu_factor+    &  1.0   &          $f_{\omega}$\\
\verb+am_gradmu_factor+&  0.05  &          $f_{\mu}$ \\
\hline
\end{tabular}
}
\label{t:MESA}
\end{table}

We use the initial settings in the MESA version 8118 code at the \verb+star/test_suite/massive_rotating+ directory for Solar chemical composition, the variable \verb+mesh_delta_coeff+ was set to 0.8 for model convergence.
The kinematic shear viscosity most significantly affects the angular momentum transport, 
and therefore we use three values of this parameter for three different models.
 
Fig. \ref{f:MESA} shows the time evolution of the angular rotational velocity $\Omega$ of the He core of a 15 $M_\odot$
star calculated by MESA (left curves) with different kinematic shear viscosity factor \verb+am_nu_visc_factor+ and by BSE (right curves) with different core-envelope coupling time $\tau_c$. The curves start at the time corresponding 
to the appearance of the He core after the main sequence. At the zero main sequence, the star is assumed to be 
rigidly rotating with the angular velocity calculated from Eq. (107) in \cite{2000MNRAS.315..543H}.
The time shift between the MESA (left curves) 
and BSE (right curves) calculations is apparently due to different treatment of convection in the stellar evolution used in MESA and BSE. It is seen from Fig. \ref{f:MESA} that the evolution of the rotational velocity of the stellar core with time 
calculated by MESA can reproduce much more simplified 
treatment using the effective core-envelope coupling, as described in Section 2 and implemented in the BSE code, by
choosing the appropriate value of the kinematic viscosity coefficient $\nu$ in the MESA code. 
Thus we conclude that the use of the core-envelope coupling parameter
$\tau_c$ in the BSE models with $\tau_c=5\times10^5-10^6$~yrs is in agreement with detailed MESA calculations.

\section{Typical evolutionary channels leading to 
rapidly rotating stellar cores}

Here we provide examples of typical evolutionary tracks of binary 
stars that lead to the formation of rapidly rotating iron cores before
collapse. Tracks are grouped in 'channels' according to the evolutionary stage of
the components before the formation of the rotating core in the common envelope stage (CE).
The designation of the evolutionary stages is the same as in the BSE code (MS -- main sequence star, 
HG -- post-MS star in the Hertzsprung gap, GB -- star at the giant branch, CHeB -- core helium burning, 
EAGB -- early asymptotic giant branch, TPAGB -- thermally pulsing AGB, HeMS -- naked helium main sequence, 
HeHG -- helium star in the Hertzsprung gap). Columns in the tables show: time in million years (T),
evolutionary stages of the components, masses of the components in solar masses (M1, M2), orbital separation in solar radii (A),
radii of the components in units of the effective Roche lobe radius (R1/RL1, R2/RL2). 

\clearpage
\begin{table}
\centering
  \caption{Channel 1: First Giant Branch + MS}
\tabcolsep 1.0 mm          
{\scriptsize
\begin{tabular}{l|cc|cc|ccc}
\hline																
	T	&	Star1	&	Star2	&	M1	&	M2	&	A	&	R1/RL1	&	R2/RL2	\\
\hline	
	0.0	&	MS	&	MS	&	8.6	&	8.4	&	203	&	0.05	&	0.05	\\
	32.4	&	HG	&	MS	&	8.5	&	8.3	&	204	&	0.11	&	0.11	\\
	32.5	&	HG	&	MS	&	8.5	&	8.3	&	205	&	1.05	&	0.12	\\
	32.5	&	\bf{GB}	&	\bf{MS}	&	8.5	&	8.4	&	201	&	\bf{CE1}&	-	\\
	32.5	&	GB	&	 -    	&	11.9	&	 -    	&	 -    	&	-    	&	 -    	\\
	32.5	&	CHeB	&	 -    	&	11.9	&	 -    	&	 -    	&	 -    	&	 -    	\\
	36.3	&	EAGB	&	 -    	&	11.6	&	 -    	&	 -    	&	 -    	&	 -    	\\
&\multicolumn{3}{l}		{Core collapse}\\
\hline	
\hline
\end{tabular}
}
\label{tab:track1}
\end{table}
\begin{table}
\centering
  \caption{Channel 2: Core He-burning + shell He-burning	}
\tabcolsep 1.0 mm          
{\scriptsize
\begin{tabular}{l|cc|cc|ccc}
\hline																
	T	&	Star1	&	Star2	&	M1	&	M2	&	A	&	R1/RL1	&	R2/RL2	\\
\hline	
	0.0	&	MS	&	MS	&	10.3	&	10.0	&	1147	&	0.01	&	0.01	\\
	23.4	&	HG	&	MS	&	10.1	&	10.0	&	1159	&	0.02	&	0.02	\\
	23.5	&	GB	&	MS	&	10.1	&	10.0	&	1160	&	0.45	&	0.02	\\
	23.5	&	CHeB	&	MS	&	10.1	&	10.0	&	1103	&	0.96	&	0.03	\\
	24.2	&	CHeB	&	HG	&	9.9	&	10.0	&	1106	&	0.66	&	0.02	\\
	24.3	&	CHeB	&	GB	&	9.9	&	10.0	&	1106	&	0.65	&	0.46	\\
	26.2	&	EAGB	&	CHeB	&	9.7	&	9.6	&	1064	&	0.97	&	0.12	\\
	26.2	&	EAGB	&	CHeB	&	9.7	&	9.6	&	1031	&	1.02	&	0.13	\\
	26.2	& \bf{EAGB}	& \bf{CHeB}	&	9.7	&	9.6	& 	1031  	& \bf{CE1}	&	-	\\
	26.2	&	CHeB	&	 -    	&	16.6	&	 -    	&	 -    	&	 	&	 -\\
	26.4	&	EAGB	&	 -    	&	15.0	&	 -    	&	 -    	&	 -    	&	 - \\
&\multicolumn{3}{l}		{Core collapse}\\
\hline
\hline
\end{tabular}
}
\label{tab:track2}
\end{table}
\begin{table}
\centering
  \caption{Channel 3: naked helium main sequence + shell H-burning}
\tabcolsep 1.0 mm          
{\scriptsize
\begin{tabular}{l|cc|cc|ccc}
\hline																
T	&	Star1	&	Star2	&	M1	&	M2	&	A	&	R1/RL1	&	R2/RL2	\\
\hline	
0.0	&	MS	&	MS	&	10.1	&	9.3	&	390	&	0.03	&	0.03	\\	
23.8	&	HG	&	MS	&	10.0	&	9.3	&	393	&	0.06	&	0.05	\\	
23.9	&	HG	&	MS	&	10.0	&	9.3	&	394	&	1.03	&	0.05	\\	
23.9	&	GB	&	MS	&	10.0	&	9.3	&	381	&	\bf{CE1}	&	-	\\	
27.7	&	HeMS	&	HG	&	2.1	&	9.2	&	52.5	&	0.03	&	0.33	\\	
27.7	& HeMS		&	HG	&	2.1	&	9.2	&	52.7	&	0.03	&	1.00	\\	
27.7	& \bf{HeMS}	&	\bf{HG}	&	2.1	&	9.2	&	52.7	&	-	&	\bf{CE2}	\\	
27.7	&	-	&	CHeB	&	-	&	8.1	&	-	&	-	&	 -	\\	
29.9	&	-	&	EAGB	&	-	&	8.0	&	-	&	-	&	-	\\	
30.0	&	-	&	TPAGB	&	-	&	7.9	&	-	&	-	&	-	\\	
&\multicolumn{3}{l}		{Core collapse}\\
\hline																
\hline
\end{tabular}
}
\label{tab:track3}
\end{table}
\begin{table}
\centering
  \caption{Channel 4: naked helium sub giant + shell H-burning}
\tabcolsep 1.0 mm          
{\scriptsize
\begin{tabular}{l|cc|cc|ccc}
\hline																
	T	&	Star1	&	Star2	&	M1	&	M2	&	A	&	R1/RL1	&	R2/RL2	\\
\hline	
0.0	&	MS	&	MS	&	11.5	&	10.7	&	163	&	0.07	&	0.07	\\
19.3	&	HG	&	MS	&	11.3	&	10.6	&	165	&	0.16	&	0.14	\\
19.3	&	HG	&	MS	&	11.3	&	10.6	&	165	&	1.01	&	0.14	\\
19.3	&	GB	&	MS	&	10.6	&	11.3	&	143	&	\bf{CE1}	&	-	\\
19.6	&	HeMS	&	MS	&	2.6	&	11.3	&	24.8	&	0.07	&	0.69	\\
22.4	&	HeHG	&	MS	&	2.4	&	11.3	&	25.8	&	0.06	&	0.94	\\
22.5	&	HeHG	&	HG	&	2.4	&	11.3	&	25.9	&	0.12	&	0.77	\\
22.5	&	HeHG	&	HG	&	2.4	&	11.3	&	25.9	&	0.13	&	1.05	\\
22.5	&	\bf{HeHG}&	\bf{HG}	&	2.4	&	11.3	&	25.9	&	-	&	\bf{CE2}	\\
22.5	&	-	&	CHeB	&	-	&	13.5	&	-	&	-	&	- 	\\
23.5	&	-	&	EAGB	&	-	&	10.8	&	-	&	-	&	-	\\
&\multicolumn{3}{l}		{Core collapse}\\
\hline
\hline
\end{tabular}
}
\label{tab:track4}
\end{table}
\begin{table}
\centering
  \caption{Channel 5: naked helium sub giant + naked helium sub giant}
\tabcolsep 1.0 mm          
{\scriptsize
\begin{tabular}{l|cc|cc|ccc}
\hline																
	T	&	Star1	&	Star2	&	M1	&	M2	&	A	&	R1/RL1	&	R2/RL2	\\
\hline		
0.0	&	MS	&	MS	&	12.5	&	12.5	&	1175.0	&	0.01	&	0.01	\\
16.7	&	HG	&	MS	&	12.3	&	12.3	&	1194.4	&	0.02	&	0.03	\\
16.7	&	GB	&	MS	&	12.3	&	12.3	&	1194.8	&	0.63	&	0.03	\\
16.7	&	GB	&	MS	&	12.3	&	12.3	&	1140.4	&	1.00	&	0.03	\\
16.7	&	GB	&	MS	&	12.3	&	12.3	&	1140.4	&	\bf{CE1}	&	-	\\
16.7	&	HeMS	&	MS	&	2.9	&	12.3	&	202.2	&	0.01	&	0.12	\\
16.7	&	HeMS	&	HG	&	2.9	&	12.3	&	202.2	&	0.01	&	0.11	\\
16.8	&	HeMS	&	HG	&	2.9	&	12.3	&	202.3	&	0.01	&	1.05	\\
16.8	& 	HeMS	&	 HG	&	2.9	&	12.3	&	202.3	&	-	&	\bf{CE2}	\\
16.8	&	HeMS	&	HeMS	&	2.9	&	2.9	&	1.8	&	0.63	&	0.62	\\
19.3	& HeHG		&	HeHG	&	2.7	&	2.7	&	1.6	&	1.00	&	0.68	\\
19.3	& \bf{HeHG}	&\bf{HeHG}	&	2.7	&	2.7	&	1.6	& \bf{CE3}	&	-	\\
19.3	&	HeHG	&	-	&	5.1	&	-	&	-	&	- 	&	-	\\
&\multicolumn{3}{l}		{Core collapse}\\
\hline																
\hline
\end{tabular}
}
\label{tab:track5}
\end{table}

\bibliographystyle{mnras}
\bibliography{ns_prog}

\end{document}